\documentclass[namedreferences]{SolarPhysics}
\usepackage[optionalrh]{spr-sola-addons} % For Solar Physics 
\usepackage{graphicx}        % For eps figures, newer & more powerfull
\usepackage{color}           % For color text: \color command
\usepackage{url}             % For breaking URLs easily trough lines
\newcommand{\etal}{{\it et al.}}

\newcommand{\aap}{    {\it Astron. Astrophys.}}
\newcommand{\apj}{    {\it Astrophys. J.}}
\newcommand{\apss}{   {\it Astrophys. Space Sci.}}
\newcommand{\astroph}{{\it astro-ph}}
\newcommand{\ieee}{   {\it IEEE Trans. Automat. Control}}
\newcommand{\jgr}{    {\it J. Geophys. Res.}}
\newcommand{\prl}{    {\it Phys. Rev. Lett.}}
\newcommand{\solphys}{{\it Solar Phys.}}
\newcommand{\sw}{     {\it Space Weather}}

\begin{document}

\begin{article}

\begin{opening}

\title{Statistical Models for Solar Flare Interval Distribution in Individual Active Regions}

\author{Y\=uki \surname{Kubo}}
\runningauthor{Y. Kubo}
\runningtitle{Flare Interval in Individual Active Regions}

   \institute{       Space Environment Group, National Institute of Information and Communications Technology, Tokyo, Japan \\
email: \url{kubo@nict.go.jp} \\ 
              Core Research for Evolutional Science and Technology, Japan Science and Technology Agency, Tokyo, Japan
             }

\begin{abstract}
This article discusses statistical models for solar flare interval distribution in individual active regions. We analyzed solar flare data in 55 active regions that are listed in the Geosynchronous Operational Environmental Satellite (GOES) soft X-ray flare catalog for the years from 1981 to 2005. We discuss some problems with a conventional procedure to derive probability density functions from any data set and propose a new procedure, which uses the maximum likelihood method and Akaike Information Criterion (AIC) to objectively compare some competing probability density functions. Previous studies on solar flare interval distribution in individual active regions only dealt with constant or time-dependent Poisson process models, and no other models were discussed. We examine three models---exponential, lognormal, and inverse Gaussian---as competing models for probability density functions in this study. We found that lognormal and inverse Gaussian models are more likely models than the exponential model for solar flare interval distribution in individual active regions. Possible mechanisms of solar flare for the distribution models are briefly mentioned. We also investigated a time-dependence of probability density functions of solar flare interval distribution briefly, and found that some active regions show time-dependence for lognormal and inverse Gaussian distribution functions. The results suggest that solar flares do not occur randomly in time; rather, solar flare intervals appear to be regulated by solar flare mechanisms. Determining of a solar flare interval distribution is an essential step in probabilistic solar flare forecasting method in space weather research. We briefly mention a probabilistic solar flare forecasting method as an application of a solar flare interval distribution analysis. The application of our distribution analysis to a probabilistic solar flare forecasting method is one of the main objectives of this study.
\end{abstract}

\end{opening}
%-------------------------------------------------

\section{Introduction}
Solar flares are one of the most energetic phenomena to occur on the solar surface. However, the mechanism of flares is still not fully understood despite many efforts to study them. Magnetic reconnection is thought to play a central role in a flare energy release. However, the trigger of the magnetic reconnection as well as the process of energy storage in active regions are also unknown. Many studies have concentrated on physical processes in an individual event to investigate the flare mechanism. Recently, new solar observational satellites such as {\it Hinode} or {\it STEREO} have been launched. Observation data from the new satellites will certainly provide a lot of new information about flare mechanisms.\par
On the contrary, the sun and solar radiations, {\it e.g.} X-ray and energetic particles, have been continuously observed for more than three decades by ground-based telescopes and satellites such as the Geosynchronous Operational Environmental Satellite (GOES). These continuous observation data include information on many past solar flares, which makes it possible to statistically analyze them. Statistical analyses of flares will be able to provide some clues about flare mechanisms; for example, an analysis of flare interval distribution may provide some clues about energy storage mechanisms.\par
Many statistical studies have been done on solar flare interval distribution. An early study by Pearce, Rowe, and Yeung (1993) investigated the flare interval distribution of hard X-ray solar flares observed by the Hard X-Ray Burst Spectrometer (HXRBS) on the Solar Maximum Mission satellite (SMM) from 1980 to 1989. They tried to fit the flare interval distribution by a sum of Poisson processes with a yearly constant mean flare interval but failed. They pointed out that the flare interval followed a power law distribution and concluded that flares did not occur randomly in time. Boffetta \etal\ (1999) analyzed soft X-ray flare data from 1976 to 1996, which are included in the GOES soft X-ray flare catalog. They found that flare intervals clearly followed a power law distribution in the range larger than six hours. They pointed out that a flare mechanism included strong correlations between consecutive flares. A similar result was found by Lepreti, Carbone, and Veltri (2001), who analyzed soft X-ray flare data from 1975 to 1999 in the GOES soft X-ray flare catalog. They argued that flare intervals are not a time-varying Poisson process, and they found that flare intervals followed a L\'evy distribution, whose tail displayed an asymptotic power law. They pointed out that flare intervals following a L\'evy distribution represented the evidence of strong correlation between consecutive flares. Wheatland (2000a) also analyzed soft X-ray flare data from 1975 to 1999 in the GOES soft X-ray flare catalog and found that flare intervals followed a power law distribution in the range larger than 10 hours. This is the same result obtained in the other studies mentioned; however, their interpretation was completely different from the former ones. They interpreted that the power law distribution could be reproduced as the sum of Poisson processes with a time-varying mean flare interval, referred to as the piecewise constant Poisson process. This idea was mentioned in Pearce, Rowe, and Yeung (1993), although they failed with this approach. Wheatland (2000a) used the mean flare interval determined by the Bayesian block procedure by Scargle (1998) instead of the yearly constant mean flare interval which was used by Pearce, Rowe, and Yeung (1993) and found that the mean flare interval distribution was approximated as an exponential function. This model was able to explain well the power law distribution of flare intervals. Wheatland (2000a) concluded that there is no evidence for strong correlation between consecutive flares.
The models introduced above focused on all flares that occurred on the entire solar disk.\par

Some authors have investigated solar flare interval distributions in individual active regions. Boffetta \etal\ (1999) analyzed soft X-ray flare data included in the GOES catalog from 1976 to 1996. They found that flare intervals in individual active regions followed a power law distribution. However, this power law distribution may not reflect a flare interval distribution in an individual active region because their data set was composed of large numbers of active region data that included quiet regions as well as very active regions. Six individual active regions, which appeared during the solar maximum of 1989 to 1991, were analyzed by Moon \etal\ (2001) based on the GOES soft X-ray flare catalog. They concluded that flare intervals in six active regions followed a single exponential distribution. Wheatland (2001a) also investigated a large number of active regions included in the GOES soft X-ray flare catalog from 1981 to 1999. They dealt with the same piecewise constant Poisson process as Wheatland (2000a) and concluded that this process was a good model for a flare interval distribution in individual active regions. Wheatland (2001a) confirmed that the constant Poisson process was a good model for six active regions that were studied by Moon \etal\ (2001). Wheatland and Litvinenko (2002) also proposed a time-dependent Poisson process model as a basic model for solar flare interval distribution. Although the piecewise constant Poisson process model by Wheatland (2000a) and Wheatland (2001a) is an approximate model of the time-dependent Poisson process model, the model can explain observational characteristics of solar flare interval distribution fairly well. To summarize these previous studies, there is only one statistical model for flare interval distribution in individual active regions. That is the time-dependent Poisson process model by Wheatland and Litvinenko (2002) or its approximate model by Wheatland (2001a), which includes the model by Moon \etal\ (2001). However, other probability density functions may also be good models because no one has investigated any other models so far.\par

Determining a more likely probability density function for solar flare intervals is one of the most essential in probabilistic solar flare forecasting. Generally, there are two approaches for probabilistic solar flare forecasting. In one approach, the locations of flares are not distinguished. The sun is therefore regarded as one flaring system. Because the flaring rate, which is characterizes a flare interval distribution, of the whole sun varies in time due to fluctuating solar activity, a time-varying flaring rate must be used in this approach. However, determining the time-dependence of flaring rate is difficult. Wheatland (2005) overcame this difficulty by making use of the Bayesian block procedure by Scargle (1998) and a power law solar flare size distribution, and constructed a probabilistic solar flare forecasting method with this approach. The method outputs a very good flare occurrence probability although it uses only the GOES data. In the other approach, the sun is regarded as a complex of many active regions. This assumption is justified because almost all flares occur in active regions, and the active regions may be independent of each other. The flaring rate of the whole sun is determined by the number of active regions and by the flaring rate of an individual active region. In this approach, it is essential to determine the flare interval distribution in individual active regions. As the first step in developing a probabilistic solar flare forecasting model using the second approach, we investigate again the flare interval distribution in individual active regions.\par

This article is organized into five sections. Section 2 describes the statistical method. In that section, a procedure to determine a probability density function by making use of the maximum likelihood method and Akaike Information Criterion (AIC) is explained. This procedure was applied to solar flare interval data, which is described in Section 3, along with some new results. Some discussions including about probabilistic solar flare forecasting method are described in Section 4. Section 5 is the conclusion.\par

\section{Statistical Method}
\subsection{Procedure to Determine Probability Density Function}
\label{sub:procedure}
The conventional procedure to determine a probability density function from any kinds of data set is as follows.
First, the data set is divided into bins, and the probability density is obtained by making use of the binned data set. We refer to this here as an empirical probability density. Then, one can infer a probability density function to fit the empirical probability density and estimate parameter values by the least-squares method. Finally, a statistical hypothesis testing of fit, like the Kolmogorov-Smirnov test or the $\chi^2$ test, is carried out to reject a null hypothesis, which states that the empirical probability density is equal to the inferred probability density function with the estimated parameter values. If the null hypothesis is not rejected, in other words, the p-value of the statistical hypothesis test is high, one may conclude that the probability density function is the inferred one with the estimated parameter values. However, rigorously speaking, it is not obvious whether the conclusion is correct or not, as discussed below.\par
This procedure has some problems. One problem concerns data binning. Parameter values that are estimated by the least-squares method can vary depending on how the data set is binned, especially bins which include the very small numbers of data strongly affect the parameter values if the data set is consists of very few data. This is attributed to a loss of information due to data binning. Another problem is in relation to statistical hypothesis testing. Because a statistical hypothesis test is essentially carried out in order to reject a null hypothesis, a null hypothesis can only be rejected and never be accepted. A result of hypothesis test, which is that a null hypothesis is not rejected, never affirmatively state that the null hypothesis is correct.\par
Taking these problems into consideration, we determine a probability density function using the following new procedure. First, the data set is divided into bins, and the empirical probability density is obtained using the binned data set. It is better to carry it out by more than one manner of binning, for example, number of bins and whether they are linearly or logarithmically binned. To prevent loss of information, we use the empirical probability density to infer probability density functions. One can infer more than one probability density function for the empirical probability density to compare with each other. Then, we estimate parameter values for all inferred probability density functions using the maximum likelihood method, which does not give rise to a loss of information (see Section \ref{sub:mlm}). Finally, we compare all probability density functions to select the most likely model. An objective comparison is carried out by using AIC (see Section \ref{sub:AIC}) although the other model comparison methods exist. AIC can be calculated very easily and it is one of the standard model selection methods in statistics, when parameter values are estimated by using the maximum likelihood method. AIC is very widely and practically used in model selection problems not only in natural science but also in social sciences, and it is a reliable method for model selection problems. These are reasons why we use AIC as model comparison method.\par
Most statistical models are approximate models because the model is inferred by a finite number of samples, that is, the statistical models do not always reflect an underlying nature of data exactly. However, since more likely statistical models can provide clues about the underlying nature of the data and good predictions for future steps, selecting a more likely model from several competing models is essentially important for statistical model identification.\par

\subsection{Maximum Likelihood Method}
\label{sub:mlm}
To provide a better understanding of the next subsection, we briefly introduce here the maximum likelihood method, although it is common in statistics. Consider that random variables $X_1,X_2,\cdots,X_n$ are distributed into a joint probability density function $p(x_1,x_2,\cdots,x_n;\mbox{\boldmath $\theta$})$, where $\mbox{\boldmath $\theta$}$ is a parameter vector we want to estimate. When samples from the joint probability density function $\mbox{\boldmath $x$}=(x_1,x_2,\cdots,x_n)$ are regarded as fixed values, likelihood function $L(\mbox{\boldmath $\theta$}|\mbox{\boldmath $x$})$ as a function of $\mbox{\boldmath $\theta$}$ is defined as
\begin{equation}
	L(\mbox{\boldmath $\theta$}|\mbox{\boldmath $x$})=p(\mbox{\boldmath $x$;$\theta$}).
	\label{likelihood_joint}
\end{equation}
If random variables $X_1,X_2,\cdots,X_n$ are independently and identically distributed into the probability density function $p(x;\mbox{\boldmath $\theta$})$, Equation (\ref{likelihood_joint}) results in
\begin{equation}
	L(\mbox{\boldmath $\theta$}|\mbox{\boldmath $x$})=\prod_{i=1}^n p(x_i;\mbox{\boldmath $\theta$}).
	\label{likelihood}
\end{equation}
The parameter $\mbox{\boldmath $\theta$}$ is estimated by maximizing the likelihood function $L(\mbox{\boldmath $\theta$}|\mbox{\boldmath $x$})$. In this case, log-likelihood ${\rm LL}(\mbox{\boldmath $\theta$}|\mbox{\boldmath $x$})$, which can be mathematically calculated easier than a likelihood, is defined as
\begin{equation}
	{\rm LL}(\mbox{\boldmath $\theta$}|\mbox{\boldmath $x$})=\log L(\mbox{\boldmath $\theta$}|\mbox{\boldmath $x$})=\sum_{i=1}^n\log p(x_i;\mbox{\boldmath $\theta$}).
	\label{loglikelihood}
\end{equation}
The log-likelihood can be maximized at $\mbox{\boldmath $\theta$=$\hat \theta$}$, namely
\begin{equation}
	\frac{\partial {\rm LL}(\mbox{\boldmath $\theta$}|\mbox{\boldmath $x$})}{\partial \theta_j}\Biggr|_{\mbox{\boldmath $\hat\theta$}}=0 \quad (j=1,\cdots,l),
	\label{LLpartial}
\end{equation}
where $l$ is the number of parameters.
The $\mbox{\boldmath $\hat \theta$}$ represents the estimated parameter values of $\mbox{\boldmath $\theta$}$, which is called a maximum likelihood estimator.
This method does not give rise to an information loss because all sample data are used to estimate parameter values. In other words, there are no uncertainties in parameter values due to data binning.

\subsection{AIC}
\label{sub:AIC}
AIC, which stands for the Akaike Information Criterion, is a criterion to select a more likely model among several competing models, and it provides an objective procedure for statistical model selection (Akaike, 1974). It is widely and practically used in model selection problems in many fields. AIC is defined very simply as
\begin{equation}
	{\rm AIC}=-2{\rm LL}(\mbox{\boldmath $\hat \theta$}|\mbox{\boldmath $x$})+2k,
	\label{AICdef}
\end{equation}
where $k$ is the number of free parameters in the model. In this definition, ${\rm LL}(\mbox{\boldmath $\hat \theta$}|\mbox{\boldmath $x$})$ is a maximum log-likelihood, Equation (\ref{loglikelihood}) with maximum likelihood estimator $\mbox{\boldmath $\hat \theta$}$. AIC divided by $-2$ is an unbiased estimator of mean log-likelihood and in relation to information entropy. According to information theory, a model that has the largest mean log-likelihood is the most likely model of several competing models; namely, the minimum AIC model is the most likely model among several competing models.\par
Some points regarding AIC need to be taken into consideration. 1) AIC is not a criterion to select a true model, but a criterion to select a more likely model. 2) Only the relative difference in AIC among several competing models is significant, and there is no significance in the absolute value of AIC. 3) A relative difference of more than one or two is a significant difference. 4) A model that has a small AIC value relative to other models is a more likely model. 5) The number of free parameters in the model must be less than $2\sqrt n$, where $n$ is the number of samples. When there are several competing models for the same data set we can select the minimum AIC model as the most likely model among these competing models. AIC provides a versatile procedure for statistical model identification.

\begin{figure}[t]
	\includegraphics{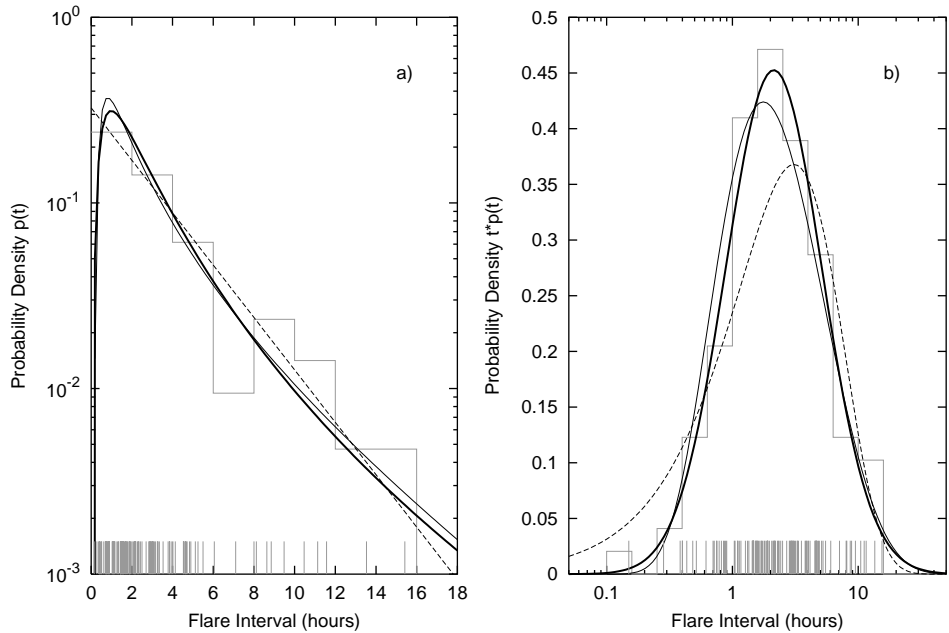}
	\caption{Probability density functions for active region NOAA5395. The gray ticks represent samples used to obtain an empirical probability density. The gray stepped line represents an empirical probability density. The dashed, thick, and thin black lines represent exponential, lognormal, and inverse Gaussian distributions, respectively. Panel (a) depicts a linearly binned data set. Panel (b) depicts a logarithmically binned data set.}
	\label{5395bin}
\end{figure}

\section{Solar Flare Interval Distribution}
\label{sec:flare_interval}
In this study we use the GOES soft X-ray flare catalog. The catalog is available on the website, \url{http://www.ngdc.noaa.gov/stp/SOLAR/ftpsolarflares.html}. An advantage in using the GOES flare catalog is that it covers a very long observation period with few interruptions. We have used the data for the years from 1981 to 2005. This period covers the maximum phase of the 21st solar cycle to the descending phase of the 23rd solar cycle. During this period, 7987 active regions were labeled, and 3161 active regions out of those 7987 (39.6\%) produced at least one flare greater than or equal to the C1.0 class of soft X-ray classification. The total number of flares greater than or equal to the C1.0 class was 40,552, and of those, 24,917 flares (61.4\%) were identified with active regions. We chose the threshold level C1.0 X-ray flux because this flux level was the typical background level in the maximum phase of solar cycle (Hudson, 1991). We define flare interval as the difference between start times of two consecutive flares produced in the same active region.\par
As an analysis example, we analyzed a data set for the active region NOAA5395. The first flare in the region occurred on 5 March 1989. After this flare, the region produced a lot of flares including the X15.0 class. The total number of flares greater than or equal to the C1.0 class was 107, and there were 106 flare intervals in that active region. In analyzing the data set using a new procedure to determine the probability density function, which we described in Section \ref{sub:procedure}, we constructed a statistical model of a probability density function for flare intervals in active region NOAA5395.\par

\begin{table}[t]
	\begin{tabular}{ccccc}
	\hline
	NOAA5395 & $\hat\mu,\hat m$ & $\hat\sigma^2,\hat\alpha^2$ & Mean & AIC \\
	\hline
	Exponential & 3.0769 & -      & 3.0769 & 452.3 \\
	\hline
	Lognormal   & 0.7561 & 0.7770 & 3.1410 & 438.4 \\
	\hline
	Inverse Gaussian & 3.0769 & 1.1795 & 3.0769 & 443.6 \\
	\hline
	\end{tabular}
	\caption{Parameter estimation results for active region NOAA5395.}
	\label{5395tbl}
\end{table}

First, we divided the 106 samples into bins, using both linear and logarithmic binning. Figure \ref{5395bin} shows an empirical probability density of flare intervals for the active region. The many gray ticks in the figure represent samples used to obtain the empirical probability density. The gray stepped line represents an empirical probability density. Figure \ref{5395bin}a plots the empirical probability density of the linearly binned samples. The ordinate axis in the figure is the fractional number of samples per bin width, which is drawn on a logarithmic scale. As the figure shows, the empirical probability density can be approximated by a straight line (dashed black line) on the log-linear coordinate. One can infer from this fact that the flare intervals are distributed exponentially. Figure \ref{5395bin}b is the empirical probability density made of logarithmically binned samples. The ordinate axis is the fractional number of samples per bin width multiplied by median values representing each bin. The two ordinates are different because both graphs were drawn to satisfy the equation $p(t)dt=tp(t)d(\log t)$. It is very interesting that the forms of these two empirical probability densities are completely different. Is it apparent from Figure \ref{5395bin}b that the exponential distribution is a good model for an empirical probability density? If anything can be inferred from this, it is that a lognormal distribution (thick black line) or an inverse Gaussian distribution (thin black line), for example, are better models than the exponential distribution. We will deal with three distribution models---exponential, lognormal, and inverse Gaussian---as competing probability density function models. The exponential, lognormal, and inverse Gaussian models are expressed respectively as 
\begin{equation}
	p(t;\mu)=\frac{1}{\mu}\exp\left(-\frac{t}{\mu}\right),
\end{equation}
\begin{equation}
	p(t;m,\sigma^2)=\frac{1}{\sqrt{2\pi\sigma^2}t}\exp\left\{-\frac{(\log t-m)^2}{2\sigma^2}\right\},
\end{equation}
and
\begin{equation}
	p(t;\mu,\alpha^2)=\sqrt{\frac{\mu}{2\pi\alpha^2t^3}}\exp\left\{-\frac{(t-\mu)^2}{2\mu\alpha^2t}\right\},
\end{equation}
where $\mu$, $m$, $\sigma^2$, and $\alpha^2$ are scalar parameters. The other models are also possible as competing probability density functions. The reason of an our selection of models are that these models can be statistically and physically interpreted as to some mechanisms related to solar flare (see Section \ref{sec:discussion}).\par

\begin{figure}[t]
	\includegraphics{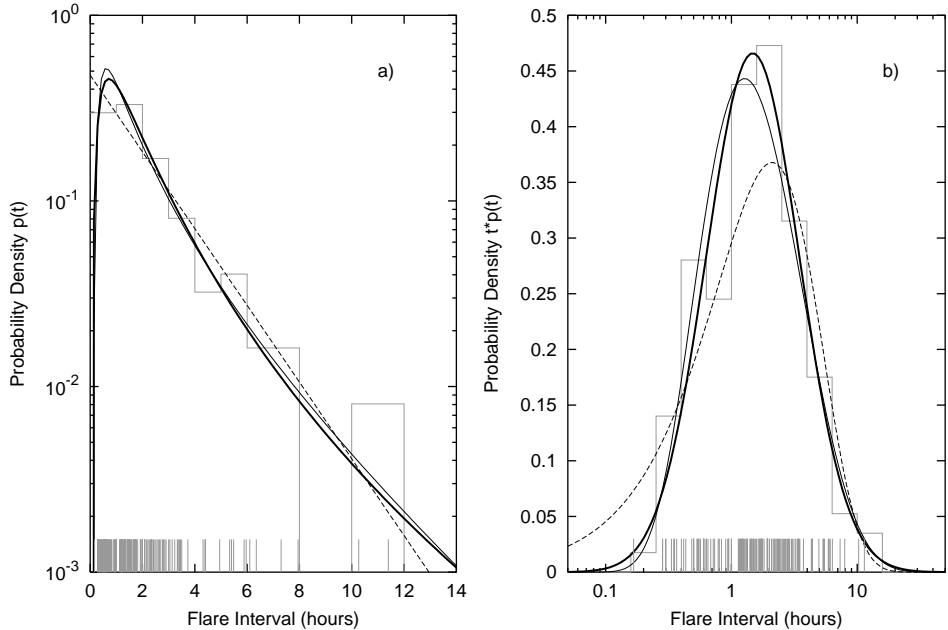}
	\caption{Same as figure \ref{5395bin}. Data set is for active region NOAA10656.}
	\label{10656bin}
\end{figure}

In the next step to determine a probability density function, we estimate parameter values using the maximum likelihood method and calculate the AIC of the three models. Solving Equation (\ref{LLpartial}) with the 106 flare interval data for the three competing probability density function models, we can estimate the maximum likelihood estimators $\hat\mu,\hat m,\hat\sigma^2$, and $\hat\alpha^2$. AIC is calculated by Equation (\ref{AICdef}). The results are summarized in Table \ref{5395tbl}. In the table, mean stands for the first moment of the distribution. In these results, although the mean values of the distributions are almost the same, there are significant differences in AIC. It is smallest for the lognormal distribution model and largest for the exponential distribution model. These results indicate that the lognormal distribution model is the most likely, and the exponential distribution model is the least likely probability density function model for this active region. These results are also shown in Figure \ref{5395bin}. The dashed, thick, and thin black lines in Figure \ref{5395bin}a and \ref{5395bin}b show exponential, lognormal, and inverse Gaussian distribution models with the maximum likelihood estimator, respectively. The lognormal distribution model fits the empirical probability density well. However, the exponential distribution model deviates from the empirical one, as can be clearly seen in Figure \ref{5395bin}b.\par

\begin{table}[t]
	\begin{tabular}{ccccc}
	\hline
	NOAA10656 & $\hat\mu,\hat m$ & $\hat\sigma^2,\hat\alpha^2$ & Mean & AIC \\
	\hline
	Exponential & 2.1008 & -      & 2.1008 & 434.1 \\
	\hline
	Lognormal   & 0.3904 & 0.7330 & 2.1315 & 414.2 \\
	\hline
	Inverse Gaussian & 2.1008 & 1.0464 & 2.1008 & 414.7 \\
	\hline
	\end{tabular}
	\caption{Results of parameter estimation for NOAA10656.}
	\label{10656tbl}
\end{table}

Another analysis example is given for active region NOAA10656. This region produced flares more frequently than any other active region during the period of analysis. The total number of flares greater than or equal to the C1.0 class was 125. We analyzed the data set to determine a more likely model of flare interval distribution in the active region. The same models as in the analysis of NOAA5395 were used for probability density functions. The results are summarized in Table \ref{10656tbl}. These results are similar to those for active region NOAA5395. Some significant differences in AIC occur, although the mean values of the distributions are almost the same for the three models. The exponential distribution model is the least likely model again. However, there is no significant difference between the lognormal and inverse Gaussian distribution models. Figure \ref{10656bin} shows the empirical probability density and probability density function models with the maximum likelihood estimator for region NOAA10656. The many gray ticks in the figure represent samples used to obtain the empirical probability density. The gray stepped line represents the empirical probability density. The dashed, thick, and thin black lines represent exponential, lognormal, and inverse Gaussian distribution models, respectively. While both lognormal and inverse Gaussian distribution models fit the empirical probability density well, the exponential distribution model again deviates from the empirical one, especially at relatively short flare intervals. This aspect is consistent with the AIC results.\par

The results of these two analyses imply that lognormal and inverse Gaussian distributions are statistically more likely models than exponential distribution for a flare interval distribution model in an individual active region. To investigate the implications of this, we analyzed data sets for 55 active regions that produced at least 51 flares greater than or equal to C1.0 class flares. The results are summarized in Table \ref{AICtbl}. The table presents three comparisons between pairs of models. Table \ref{AICtbl}a shows a comparison between lognormal and exponential distribution models. Out of the 55 active regions, it is indicated in 40 of them (72.3\%) that the lognormal distribution model is more likely than the exponential one, and only 8 active regions (14.5\%) are the opposite. There are no significant difference in 7 of the active regions (12.7\%). These results indicate that the lognormal distribution model is more likely than the exponential one for an individual active region. Table \ref{AICtbl}b compares the inverse Gaussian and exponential distribution models. Here, in 31 active regions (56.4\%) out of 55, the inverse Gaussian distribution model is indicated to be more likely than the exponential one, and 18 active regions (32.7\%) are the opposite. No significant difference was found in 6 of the active regions (10.9\%). The results indicate that the inverse Gaussian distribution model is more likely than the exponential one for an individual active region, although the difference is not very large. Table \ref{AICtbl}c shows a comparison between the lognormal and inverse Gaussian distribution models. In 30 active regions (54.5\%) out of 55, the lognormal distribution model is more likely than the inverse Gaussian one, and only 4 active regions (7.3\%) indicate the opposite. There was no significant difference in 21 of the active regions (38.2\%). These results indicate that the lognormal distribution model is more likely than the inverse Gaussian model; however, since no significant difference was found in more than a third of the 55 active regions, the difference is minor. To summarize these results, the lognormal distribution model is statistically the most likely, and the exponential distribution model is statistically the least likely of these three models for an individual active region.\par

\begin{table}[t]
	\begin{tabular}{cccc}
	\multicolumn{4}{l}{a) Lognormal - Exponential} \\
	\hline
	Lognormal        & Exponential      & No difference & Total      \\
	\hline
	40 (72.3\%)      & 8 (14.5\%)       & 7 (12.7\%)    & 55 (100\%) \\
	\hline
	\multicolumn{4}{l}{} \\
	\multicolumn{4}{l}{b) Inverse Gaussian - Exponential} \\
	\hline
	Inverse Gaussian & Exponential      & No difference & Total      \\
	\hline
	31 (56.4\%)      & 18 (32.7\%)      & 6 (10.9\%)    & 55 (100\%) \\
	\hline
	\multicolumn{4}{l}{} \\
	\multicolumn{4}{l}{c) Lognormal - Inverse Gaussian} \\
	\hline
	Lognormal        & Inverse Gaussian & No difference & Total      \\
	\hline
	30 (54.5\%)      & 4 (7.3\%)        & 21 (38.2\%)   & 55 (100\%) \\
	\hline
	\end{tabular}
	\caption{Comparison by AIC of competing model pairs for 55 active regions.}
	\label{AICtbl}
\end{table}

\section{Discussion}
\label{sec:discussion}
We consider the possible origin of the three distributions---exponential, lognormal, and inverse Gaussian---for solar flare interval distribution. Exponential distribution is related to a random process. If the flare occurrence probability per unit time interval at time $t$ does not depend on the time history of the occurrence and it is constant in time, a flare occurs randomly in time. This process is called a Poisson process. Time intervals of two consecutive events in a Poisson process are distributed exponentially. This fact means that if a flare interval follows an exponential distribution, the flare occurrence is regulated only by randomness itself. It might be interpreted that there is no relationship between solar flare intervals and flare energy storage mechanisms. Lognormal distribution is related to the central limit theorem in statistics. This theorem states that when random variables $X_1,X_2,\cdots,X_n$ are independently distributed into any kind of probability density function with finite variance, the probability density function of random variable $X=X_1+X_2+\cdots+X_n$ approaches a normal distribution with $n\rightarrow\infty$. One can recognize the following from the theorem. When random variables $Y_1,Y_2,\cdots,Y_n$ are independently distributed into any kind of probability density function with finite variance, the probability density function of random variable $Y=Y_1\cdot Y_2\cdots Y_n$ approaches a lognormal distribution with $n\rightarrow\infty$ because $\log Y=\log Y_1+\log Y_2+\cdots+\log Y_n$. We can infer from the theory that if the flare intervals follow a lognormal distribution they can be regulated as the product of many variables. This is a statistical interpretation of lognormal distribution of flare intervals. However, a physical interpretation of the distribution is unknown. Inverse Gaussian distribution is related to a kind of diffusion process. One can consider the diffusion process $dX(t)=\lambda dt+\delta dW(t)$, where $dX(t)$, $dt$, and $dW(t)$ are the state variable, time, and Wiener process, the same as in Gaussian noise, respectively. The first term on the right side of the equation is a continuously changing term, and the second term on the right side is a randomly changing term. The $\lambda$ is the average changing rate of a state variable, and $\delta^2$ is a diffusion coefficient. Time $t$, when $X$ attains $X_f$ starting with $X=X_0$ at $t=0$, is distributed as an inverse Gaussian. This indicates that if a probability density function for flare intervals is represented by an inverse Gaussian distribution model, the flare energy storage mechanism might be a kind of diffusion process, and the following flare mechanism is possible. Energy $E$ in an active region is stored by various kinds of diffusion processes from (meta)stable energy state $E_0$. When the energy attains the specific energy state $E_f$, the state of the active region will be unstable and release the energy. A flare occurs, and the energy state returns to (meta)stable energy state $E_0$. This mechanism is similar to an early model proposed by Rosner and Vaiana (1978), involving exponential energy storage followed by random energy release. In their model a correlation is expected between the amount of released energy and the flare interval because of the exponential energy storage. However, no such correlation was found in hard X-ray (Wheatland, 2000b) or soft X-ray (Moon \etal,\ 2001) observations. In the mechanism described here, such a correlation does not appear because the energy storage mechanism is a kind of diffusion process. According to this mechanism, flares do not occur randomly in time although they are independent of each other. From the viewpoint of physical interpretation of probability density functions, it is concluded that inverse Gaussian distribution is the most likely model of these three models. Taking into account these interpretations of probability density functions and the statistical result discussed in Section \ref{sec:flare_interval}, we conclude that the lognormal and inverse Gaussian distribution models are more likely than the exponential model for flare interval distribution in an individual active region.\par

In the analyses, we assumed that the parameter values of probability density functions for all active regions were constant in time. However, some active regions may show time-varying parameter values for Poisson process (Wheatland, 2001a). For these active regions, we need to deal with a time-dependent Poisson process instead of a time-independent Poisson process that corresponds to an exponential distribution model. However, we cannot compare directly a time-dependent Poisson process with the other models by AIC since it is difficult to determine a time-dependence of parameter values by using a finite number of samples. Instead of a comparison by using AIC, we test a hypothesis, which is that solar flare occurrence in an individual active region is a local Poisson process. We follow a local Poisson hypothesis test introduced by Bi, B$\ddot{\rm o}$rner, and Chu (1989). This test was applied to all flare data from 1975 to 1999 listed in the GOES soft x-ray flare catalog by Lepreti, Carbone, and Veltri (2001) and they showed that solar flare occurrence on the entire solar disk was deviate from a local Poisson process. We also apply the test to solar flare occurrence in individual active regions. The results of the test imply that a local Poisson process is not appropriate for a solar flare occurrence in individual active regions and there may be more likely models than a weighted sum of exponential distribution models for solar flare interval distributions in individual active regions. This result along with the interpretations of probability density functions and the statistical result means that solar flares do not occur randomly in time; rather, each flare occurrence might be regulated by various energy storage mechanisms.\par

For a local Poisson hypothesis test described above, we have to note that a simulation of Poisson process with an artificial selection bias to mimic a compiling procedure of the GOES flare catalog could produce a similar departure from a local Poisson process for the GOES flare catalog events (Wheatland, 2001b). This may provide an alternative interpretation of the result of the local Poisson hypothesis test for GOES flare catalog events.\par

Wheatland (2001a) showed that short flare interval events are underestimated in the GOES soft x-ray flare catalog. If this is true, the underestimation might lead to results that are slightly different from those obtained in our analysis because an obscure effect may distort a flare interval distribution. It seems that short interval events are absent in Figures \ref{5395bin} and \ref{10656bin}, and the absence might be the reason why AIC comparison prefers lognormal and inverse Gaussian to exponential distribution. To investigate the effect partly, we did same analyses described in Section \ref{sec:flare_interval} for all flare data for 55 active regions, which includes A and B class flares listed in GOES catalog. The data include a large number of short interval events than the data used in Section \ref{sec:flare_interval}. This analysis leads to somewhat better result for the lognormal and the inverse Gaussian distributions in AIC comparison of model pairs. The result implies that addition of short interval events makes AIC comparison prefer the lognormal and the inverse Gaussian distribution to the exponential distribution. This investigation shows that the absence of short interval events is not always the reason why AIC prefers the lognormal and the inverse Gaussian distribution to the exponential distribution.\par

Moon \etal\ (2001) reported and Wheatland (2001a) confirmed that exponential distribution models are good models for flare intervals of highly productive active regions NOAA5395, 5747, 6233, 6545, 6659, and 6891. We also analyzed these active region data except for NOAA6233, which produced fewer than 51 flares. As already discussed in Section \ref{sec:flare_interval}, lognormal was the most likely, and exponential was the least likely model for NOAA5395. For NOAA5747 there were no significant differences between the three models, that is, the three models are equally likely models for the region. For NOAA6545 there was no significant difference between the lognormal and inverse Gaussian models. However, the difference between these two models and the exponential model was significant. The exponential model was the least likely model of the three. For NOAA6659 the most likely model was the inverse Gaussian model. The second most likely one was the lognormal model. The least likely was the exponential model. The difference between the inverse Gaussian and lognormal models was much less than the difference between the lognormal and exponential models. For NOAA6891 there were no significant differences between the three models, that is, the three models are equally likely models for the region. These comparisons support our results, which is that lognormal and an inverse Gaussian models are more likely than an exponential model.\par

As the sun rotates on its axis, the maximum period an active region can be located on the solar disk is about 15 days, that is, the maximum flare interval in an individual active region is about 15 days. Because the data we used had a sampling interval of one minute, the minimum flare interval in an individual active region was one minute. This means that we have to use truncated probability density functions to estimate parameter values by the maximum likelihood method and to calculate AIC. However, the probability of a flare interval being longer than 15 days or shorter than 1 minute is very small. For example, the probabilities for NOAA5395 are almost zero (less than $10^{-7}$) for the three probability density functions. The parameter values are not affected by these changes. The only exception is the exponential function, in which the probability of a flare interval shorter than one minute is about $0.005$. This does slightly affect the parameter and AIC values. However, the comparison results for NOAA5395 do not change. For all of the active regions we analyzed, only 1 active region out of 55 would be affected. In this one region, the comparison result between the lognormal and exponential models would change from "lognormal" to "no difference". This does not affect the overall analysis results.\par

As already mentioned in this section, we assumed that the parameter values of probability density functions for all active regions were constant in time. However, some active regions may show time-varying parameter values for lognormal and inverse Gaussian distributions. Actually, we investigated a time-dependence of parameter values for lognormal and inverse Gaussian distributions by dividing a data for an individual active region into two halves, and determining parameter values for the lognormal and inverse Gaussian distributions for two divided data. Some active regions show that there are large differences between the parameter values for first and second half of data. This investigation implies that some active regions show time-varying parameter values for the lognormal and inverse Gaussian distributions. For these active regions, time-dependent lognormal and inverse Gaussian distribution models are more likely than time-independent models. Time-dependent lognormal and inverse Gaussian distribution models may account for some active regions for which these distributions are less likely models by the AIC comparison.\par

There are various applications for a flare interval distribution analysis. One of these is a probabilistic solar flare forecasting method. Solar flare forecasting is a very important topic in space weather research. Since a solar flare forecasting based on underlying physics is unlikely to be realized, the method has to be probabilistic. For the purpose of probabilistic flare forecasting, it is essential to determine a more likely probability density function model for flare interval distribution in an individual active region. As already discussed, short flare interval events may be obscured and the effect may distort a flare interval distribution. Although it is not certain if the obscure effect may influence a solar flare forecasting, the effect is not taken into consideration in solar flare interval analyses for probabilistic solar flare forecasting because the obscure events cannot be observed. The definition of a flare occurrence probability for active region $i$ is that at least one flare will occur within time $t$ under the condition where no flare occurred during time $T_i$ from the last flare to now. From this definition, the flare occurrence probability is expressed as $P_i(t|T_i)=1-\Phi_i(T_i+t)/\Phi_i(T_i)$, where $T_i$ is the time from the last flare in active region $i$, and $\Phi_i(t)=\int_t^\infty p_i(s)ds$, where $p_i(s)$ is a probability density function for flare interval in active region $i$. This is a conditional probability. If a probability density function is represented by an exponential, the occurrence probability is independent of the history of flare occurrence, namely, the occurrence probability is constant in time. In contrast, if a probability density function is represented by a lognormal or inverse Gaussian, the occurrence probability depends on time $T_i$. This means that the occurrence probability varies in time. As we already mentioned in the discussion, some active regions may show time-varying parameter values for solar flare interval distribution models. For these active regions, it is better to use piecewise constant probability density functions ({\it e.g.} Wheatland, 2001a for an exponential distribution function) although the piecewise constant probability density function is a convenient approximation to time-dependent distribution models. A probability of solar flare occurrence using piecewise constant probability density functions will be better than that using a time-independent probability density function, which represents a time average of probability density functions during active regions transiting the solar disk. Piecewise constant lognormal and inverse Gaussian distribution models may output the better prediction than piecewise constant Poisson model because lognormal and inverse Gaussian distribution models may be more likely than an exponential distribution model for each piece of time series. Since almost all flares occur in active regions, and active regions may be independent of each other, the sun can be regarded as a complex system composed of many active regions. Based on this viewpoint, a flare occurrence probability on the solar disk is expressed as a combination of flare occurrence probabilities of many individual active regions. While flare occurrence probabilities for all active regions on the solar disk are determined, the flare occurrence probability on the solar disk $P(t|T)$ can be expressed as $P(t|T)=1-\prod_{i=1}^n(1-P_i(t|T_i))$, where $T$ is a minimum of time $T_i$ among all active regions on the solar disk. Some difficulties still arise in obtaining concrete calculations with this probabilistic solar flare forecasting method, for example, determining a probability density function for an active region that has just emerged from the east limb of the sun. However, combined with other data, this method is expected to be used as a probabilistic solar flare forecasting method.\par

\section{Conclusion}
We analyzed solar flare data listed in the GOES soft X-ray flare catalog to estimate a more likely model of solar flare interval distributions in individual active regions. We proposed a new procedure to select a more likely model among some competing models. For an objective comparison in the procedure to determine a probability density function, we used AIC and obtained a new result, which was that lognormal distribution and inverse Gaussian distribution models are more likely models than an exponential distribution model for the probability density function of solar flare interval distribution in individual active regions. We also carried out a simple test to investigate a time-dependence of lognormal and inverse Gaussian probability density function for solar flare interval distribution, and found that some active regions show time-dependence. A hypothesis of a local Poisson process for a solar flare occurrence in individual active regions was also investigated, and we confirmed a result, which was that a local Poisson process is not appropriate for the solar flare occurrence in individual active regions. These results mean that solar flares in an individual active region do not occur randomly in time and imply that solar flare intervals are regulated by solar flare mechanisms. For example, if a probability density function is represented by an inverse Gaussian distribution model, solar flare intervals might be determined by a diffusion process. The results will provide some clues about solar flare mechanisms, especially the flare energy storage process. For practical applications of solar flare interval distribution analysis, we briefly mentioned a method to estimate the probability of solar flare occurrences. This method, along with other data, will be applicable as a probabilistic solar flare forecasting method.

%%%%%%%%%%%%%%%%%%%%%%%%%%%%%%%%%%%%%%%%%%%%%%%%%%%%%%%%%%%%%%%%%%%%%%%%%%%
\begin{acks}
We wish to thank the National Geophysical Data Center, the National Oceanic and Atmospheric Administration, U.S. Dept. of Commerce for providing information on the GOES soft x-ray flare catalog and also thank anonymous referee for useful comments to improve the manuscript.
\end{acks}

%%%%%%%%%%%%%%%%%%%%%%%%%%%%%%%%%%%%%%%%%%%%%%%%%%%%%%%%%%%%%%%%%%%%%%%%%%%

%%% BIBLIOGRAPHY %%%%%%%%%%%%%%%%%%%%%%%%%%%%%%%%%%%%%%%%%%%%%%%%%%%%%%%%%%%

\end{article} 
\end{document}